\documentclass{aa} 

\usepackage{psfig}

\def\B40{BD$+40^\circ 4124$}
\def\bd40{BD$+40^\circ 4124$}
\def\BD61{BD$+61^\circ 154$}
\def\bd61{BD$+61^\circ 154$}

\def\Ic{I$_{\rm C}$}
\def\Nk{${\cal N}_{\rm K}$} 
\def\lsim   {{_{<}\atop^{\sim}}}
\def\gsim   {{_{>}\atop^{\sim}}}

\begin{document}
\thesaurus{08(08.06.2; 08.16.5; 13.09.6)}
\title{The onset of cluster formation around Herbig Ae/Be stars}
\author{Leonardo Testi\inst{1,2}, Francesco Palla\inst{2} \and
Antonella Natta\inst{2}}
\offprints{L. Testi: lt@astro.caltech.edu}
\institute{
Division of Physics, Mathematics and Astronomy, California Institute of Technology,
MS 105-24, Pasadena CA 91125, USA
\and
Osservatorio Astrofisico di Arcetri,
Largo E. Fermi 5, I-50125 Firenze, Italy}
\date{Received xxxx; accepted xxxx}
\maketitle
\markboth{L. Testi {\it et al.}: The onset of cluster formation around Herbig Ae/Be stars}{L. Testi {\it et al.}: The onset of cluster formation around Herbig Ae/Be stars}
\begin{abstract}

The large body of near infrared observations presented in Testi et
al.~(\cite{Tea97}; \cite{Tea98}) are analysed with the aim of 
characterizing the young stellar clusters surrounding
Herbig~Ae/Be stars.
The results confirm the tendency of early Be stars to 
be surrounded by dense
clusters of lower mass ``companions'', while Ae stars are never found to
be associated with conspicuous groups. The transition between the 
different environments appears to occur smoothly from Ae to Be 
stars without a sharp threshold.

No correlation of the richness of the stellar groups detected is found
with the galactic position or the age of the central Herbig~Ae/Be star.

The stellar volume densities estimated for the groups surrounding
pre--main-sequence stars of intermediate mass show the transition from 
the low density aggregates of T Tauri stars and the dense 
clusters around massive stars.

Only the most massive stars (10-20~M$_\odot$) are found to be associated
with dense ($\sim 10^3\rm\,\,pc^{-3}$) stellar clusters. This is exactly the
mass regime at which the conventional accretion scenario for isolated star
formation faces theoretical problems. Thus our findings strongly supports
the idea that the formation of high-mass stars is influenced
by dynamical interaction in a young cluster environment.

\end{abstract}

\section{Introduction}

The tendency of stars to form in  approximately coeval groups, rather than 
in isolation, is by now a well
established observational fact
(Lada et al.~\cite{Lea93}; Zinnecker et al.~\cite{Zea93};
Gomez et al.~\cite{Gea93}; Hillenbrand~\cite{H95},~\cite{H97};
Testi et al.~\cite{Tea97}).
A result which sets stringent constraints to the
theory is the fact that the ``richness'' of the groups depends
on the mass of the most massive star in the cluster:
low-mass stars form in loose
groups of a few objects per cubic parsec (Gomez et al.~\cite{Gea93}), 
which in the following we will call ``aggregates'', while
high-mass stars are usually found to form in dense clusters with densities up
to 10$^4$ objects per cubic parsec in the case of the Trapezium 
cluster (see e.g. McCaughrean \&
Stauffer~\cite{McCS94}; Hillenbrand~\cite{H97}).  The transition
between these two modes of formation occurs in the mass interval
$2\lsim\rm M/M_\odot \lsim 15$.

Herbig Ae/Be stars (Herbig~\cite{H60}) are 
pre--main-sequence (PMS)
stars of intermediate-mass located outside complex star forming regions.
These stars are sufficiently old (age $\sim0.5-5$~Myr) to be optically
visible, but still young enough that any population of lower mass stars born
in the same environment has not had time to move away from their birthplace.
Thus, the fields around Herbig Ae/Be stars represent ideal targets to study
the transition from aggregates to dense clusters and to empirically probe the
onset of cluster formation.

Multiwalength studies of the environments of several Herbig Ae/Be stars
in the optical, near infrared and millimeter 
(LkH$\alpha$~101: Barsony et al.~\cite{Bea91}, Aspin \& Barsony~\cite{AB94};
BD$+$40$^\circ$4124: Hillenbrand et al.~\cite{Hea95}, 
Palla et al.~\cite{Pea95}; MacH12: Plazy \& M\'enard~\cite{PM97})
have shown that
the young clusters are still partially embedded in the parent molecular clouds
and that near infrared observations, especially at K-band ($2.2\,\mu$m),
are best suited to detect the less massive companions to 
the Herbig Ae/Be star itself.
Observations at near infrared wavelengths of a consistent number (16) 
of fields
around Herbig Ae/Be stars have been presented by Li et al.~(\cite{Lea94}).
Their study is focused at the detection of bright companions
and/or diffuse emission which may affect large beam photometry of the 
Herbig stars, and, for this reason, it is limited to a small area 
around each star, much smaller than the extent of the known clusters
(e.g. LkH$\alpha$~101, BD$+$40$^\circ$4124).

The first statistical study aimed at investigating the properties
of star formation around Herbig Ae/Be stars using  K-band images
has been carried out by  Hillenbrand~(\cite{H95}).
In spite of the relatively small number of observed fields (17), Hillenbrand
has found evidence for a correlation between the mass of the Herbig Ae/Be 
star and the surface density of K-band stars detected around it.
These initial results were confirmed by Testi et al.~(\cite{Tea97};
hereafter Paper~I),
who obtained J, H and K images of moderately large fields around 19 Herbig
Ae/Be stars. Using various methods to define the
richness of the star cluster, in addition to simple K-band star counts, 
in Paper~I we have shown a clear dependence of the richness of the 
embedded clusters with the spectral type of the Herbig Ae/Be star. 
Moreover, our data seem to indicate that the clustered mode of 
star formation appears at a detectable level only for stars of spectral 
type B5--B7 or earlier. Whether the transition from isolated stars to 
clusters has a threshold or is a smooth function of stellar type (or, better,
mass) is still unclear because of the small statistical significance of the
samples studied so far.

In order to complete our systematic study of Herbig stars, we have 
observed in the near infrared a second sample of 26 fields selected
from the compilations of Finkenzeller \& Mundt~(\cite{FM84}) and Th\'e et
al.~(\cite{Tea94}).
Among the observed stars, we included Z~CMa whose spectral type is F5
and whose membership to the Herbig group has been disputed. However, we will
not consider it in the analysis of the clustering properties.
Together with the 19 stars analysed in Paper~I, our final sample consists of
44 objects covering almost uniformly the whole spectral range from O9 to A7 
stars.

We have included in our sample 33 out of the 39 stars (85\%) with declination
greater than $-11.5^\circ$ listed in Finkenzeller \& Mundt~(\cite{FM84})
and replaced the remaining six objects with 11 stars of similar spectral type
taken from the updated catalog of Th\'e et al.~(\cite{Tea94}, Table I
of members and candidate members). Thus, we are confident that the final
sample gives a fairly complete representation of the whole class of Herbig
Ae/Be stars and that the inferences discussed in this paper 
should have a solid physical and statistical meaning.

In a companion paper (Testi et al.~\cite{Tea98}; hereafter Paper~II), 
we have collected the
observational material (images, colour-colour diagrams and stellar density
profiles) of each of the combined sample of 44 fields. 
In this paper, we present the results of the analysis of this large body
of observations aimed at the detection, characterization and comparison 
of the small star clusters around intermediate-mass PMS stars.

\section{Results}
\label{sres}

\subsection{Richness Indicators}

%
%
%
\begin{table*}
\caption[]{\label{trind}Values of the {\it richness indicators}.}
\vskip 0.3cm
\begin{tabular}{lcccccccccc}
\hline
Star&Type&Age (Myr)&${\cal N}_{\rm K}$&I$_{\rm C}$&${\cal N}^0_{\rm K}$&${\cal N}^2_{\rm K}$&I$^0_{\rm C}$&I$^2_{\rm C}$&Log($\rho_{{\cal N}_{\rm K}}$)&Log($\rho_{\rm I_{\rm C}}$)\\
\hline
V645~Cyg         &O7&--&$>$5 &$ 29.5\pm2$ & $>$5& $>$5&$>29.5\pm 2$&$>29.5\pm 2$&2.1& 2.9\\
MWC~297          &O9&--&37   &$ 20.4\pm1$ &   24&   23&$ 14.3\pm 1$&$ 13.3\pm 1$&3.0& 2.8\\
MWC~137          &B0&--&$>$59&$ 76.0\pm9$ &   57&   55&$ 70.1\pm 5$&$ 64.4\pm 4$&3.2& 3.4\\
R~Mon            &B0&--&0    &$-12.8\pm3$ &    0&    0&$ -5.1\pm 3$&$ -5.1\pm 3$&0.0& 0.0\\
BHJ~71           &B0&--&4    &$  4.0\pm3$ &    1&    1&$ -0.4\pm 1$&$ -0.3\pm 1$&2.0& 2.1\\
MWC~1080         &B0&--&$>$9 &$ 31.0\pm3$ & $>$9& $>$9&$>31.0\pm 3$&$>31.0\pm 3$&2.4& 3.0\\
AS~310           &B0&--&$>$37&$ 70.0\pm17$&$>$34&$>$34&$>70.2\pm17$&$>70.2\pm17$&3.0& 3.3\\
RNO~6            &B1&--&$>$11&$ 11.0\pm1$ &$>$11&$>$11&$>11.0\pm 1$&$>11.0\pm 1$&2.5& 2.5\\
HD~52721         &B2&--&10   &$ 20.5\pm4$ &    5&    5&$ 11.1\pm 8$&$ 11.1\pm 8$&2.4& 2.8\\
BD+$65^\circ$1637&B2&--&29   &$ 75.0\pm5$ &   24&   28&$ 58.4\pm 4$&$ 63.4\pm 4$&2.9& 3.3\\
HD~216629        &B2&--&29   &$ 34.0\pm6$ &   23&   22&$ 27.0\pm 6$&$ 25.3\pm 7$&2.9& 3.0\\
BD+$40^\circ$4124&B2&--&19   &$ 11.0\pm3$ &   16&   16&$ 12.6\pm 2$&$ 12.9\pm 2$&2.7& 2.5\\
HD~37490         &B3&--&9    &$  9.9\pm3$ &    6&    6&$  3.4\pm 1$&$  3.8\pm 1$&2.4& 2.5\\
HD~200775        &B3&--&8    &$  1.9\pm1$ &    7&    7&$  1.0\pm 1$&$  1.0\pm 1$&2.3& 1.8\\
MWC~300          &Be&--&$>$2 &$ 21.0\pm8$ & $>$2& $>$2&$>21.9\pm 8$&$>21.9\pm 8$&1.7& 2.8\\
RNO~1B           &Be&--&12   &$  9.7\pm1$ &   12&   12&$  8.9\pm 1$&$  9.0\pm 8$&2.5& 2.5\\
HD~259431        &B5&0.05&2    &$  0.9\pm2$ &    0&    0&$ -1.7\pm 3$&$ -0.2\pm 2$&1.7& 1.4\\
XY~Per           &B6&2.0&3    &$ 11.3\pm3$ &    2&    5&$ -0.5\pm 1$&$ -0.1\pm 1$&1.9& 2.5\\
LkH$\alpha$~25   &B7&0.2&11   &$ 14.5\pm5$ &    8&    5&$ 10.3\pm 3$&$  9.0\pm 3$&2.5& 2.6\\
HD~250550        &B7&0.5&4    &$  2.2\pm2$ &    2&    2&$  0.8\pm 1$&$  0.8\pm 1$&2.0& 1.8\\
LkH$\alpha$~215  &B7&0.1&7    &$  3.9\pm1$ &    6&    4&$  5.4\pm 1$&$  3.6\pm 1$&2.3& 2.1\\
LkH$\alpha$~257  &B8&2.0&15   &$  5.5\pm6$ &   14&   16&$  8.3\pm 3$&$  9.5\pm 4$&2.6& 2.2\\
BD+$61^\circ$154 &B8&0.1&8    &$ -1.4\pm3$ &    4&    4&$  2.0\pm 1$&$  2.4\pm 1$&2.3& 0.0\\
VY~Mon           &B8&0.05&25   &$ 23.2\pm5$ &   18&   16&$ 14.9\pm 2$&$ 12.9\pm 2$&2.8& 2.8\\
VV~Ser           &B9&1.0&24   &$ 16.9\pm5$ &   20&   23&$  3.1\pm 2$&$  3.0\pm 2$&2.8& 2.7\\
V380~Ori         &B9&1.0&3    &$ -2.0\pm2$ &    3&    3&$ -1.5\pm 3$&$ -1.5\pm 3$&1.9& 0.0\\
V1012~Ori        &B9&--&4    &$  1.9\pm2$ &    4&    4&$  1.0\pm 1$&$  1.0\pm 1$&2.0& 1.8\\
LkH$\alpha$~218  &B9&1.0&8    &$  2.0\pm5$ &    5&    7&$  1.7\pm 1$&$  2.0\pm 1$&2.3& 1.8\\
AB~Aur           &A0&1.5&$>$3 &$  3.0\pm6$ & $>$1& $>$3&$ -1.0\pm 2$&$ -2.3\pm 2$&1.9& 2.0\\
VX~Cas           &A0&1.0&13   &$  4.5\pm4$ &   34&   35&$  3.9\pm 4$&$  4.4\pm 4$&2.5& 2.1\\
HD~245185        &A2&7.0&10   &$  4.5\pm5$ &   19&   19&$  5.0\pm 1$&$  5.0\pm 1$&2.4& 2.1\\
MWC~480          &A2&7.0&$>$3 &$  5.0\pm6$ & $>$9& $>$9&$ -2.4\pm 2$&$ -2.4\pm 2$&1.9& 2.2\\
UX~Ori           &A2&1.0&0    &$ -0.3\pm1$ &    0&    0&$ -0.1\pm 1$&$ -0.1\pm 1$&0.0& 0.0\\
T~Ori            &A3&1.0&5    &$  1.0\pm2$ &    4&    7&$  0.4\pm 1$&$  1.9\pm 2$&2.1& 1.5\\
IP~Per           &A3&7.0&3    &$  5.3\pm4$ &    6&    6&$ -0.7\pm 3$&$ -0.7\pm 3$&1.9& 2.2\\
LkH$\alpha$~208  &A3&1.0&4    &$  2.2\pm5$ &    3&    2&$  1.2\pm 1$&$  2.0\pm 1$&2.0& 1.8\\
MWC~758          &A3&6.0&$>$2 &$  3.4\pm1$ & $>$3& $>$3&$ -0.3\pm 1$&$ -0.3\pm 1$&1.7& 2.0\\
RR~Tau           &A4&0.1&7    &$  0.8\pm6$ &    2&    2&$ -0.4\pm 4$&$ -3.2\pm 3$&2.3& 1.4\\
HK~Ori           &A4&7.5&7    &$  2.2\pm1$ &   11&   11&$  2.2\pm 1$&$  2.2\pm 1$&2.3& 1.8\\
Mac~H12          &A5&--&15   &$  5.1\pm1$ &   21&   21&$  5.9\pm 1$&$  6.6\pm 1$&2.6& 2.2\\
LkH$\alpha$~198  &A5&10.0&6    &$-10.6\pm11$&   14&   14&$ -9.4\pm10$&$ -8.4\pm 9$&2.2& 0.0\\
Elias~1          &A6&--&$>$2 &$  2.0\pm3$ & $>$2& $>$2&$  1.0\pm 1$&$  1.0\pm 1$&1.7& 1.8\\
BF~Ori           &A7&3.0&4    &$  1.1\pm1$ &    3&    3&$  1.1\pm 1$&$  1.1\pm 1$&2.0& 1.5\\
LkH$\alpha$~233  &A7&4.0&2    &$  1.0\pm1$ &    4&    5&$  0.6\pm 1$&$  1.5\pm 1$&1.7& 1.5\\
\hline
\end{tabular}
\end{table*}
%

\begin{figure}
\centerline{\psfig{figure=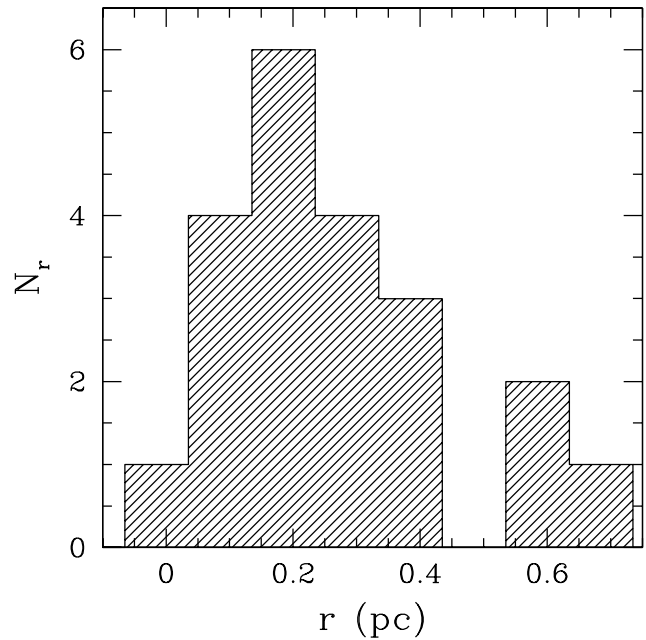,width=8.8cm}}
\caption[]{Distribution of the cluster radius of 21 stellar groups
detected by computing the source surface density profiles
obtained in Paper~II.}
\label{fsize}
\end{figure}

In our previous studies we have identified the most suitable indicators
of the richness of embedded stellar cluster that greatly reduce the problem
of background/foreground contamination. The two quantities of interest
are the number of stars in the K-band image within 0.21 pc from
the Herbig star with an absolute K magnitude M$_{\rm K}\leq 5.2$ (\Nk),
and the integral over distance of the source surface density profile
subtracted by the average source density measured at the edge of each field,
(\Ic).
The choice of a radius of 0.2 pc for computing \Nk\ is suggested by the 
typical value of the cluster size determined in Paper~II, as the 
distance from the Herbig star where the sources surface density
profile reaches the background level. 
As illustrated in Fig.~\ref{fsize}, the distribution
of the radius of the stellar density enhancement
shows a clear peak around $r\sim$0.2~pc.
This result is in good agreement
with that found in various young stellar clusters (Hillenbrand~\cite{H95};
Carpenter et al.~\cite{Car97}).
It is remarkable that stellar groups with a few to several hundred members
share similar sizes, corresponding to the typical dimensions of dense 
cores in molecular clouds.

The values of \Nk\ and \Ic\ computed for the 44 stars of our complete sample
are given in Columns~3 and 4 of Table~\ref{trind}, together with 
the name and spectral type of the Herbig Ae/Be star at the center 
of each field. The uncertainty
on \Ic\ has been calculated by propagating the error in the
determination of the background stellar density at large distances from the
Herbig star.
As in Paper~I, we discuss the dependence of the
richness indicators on the spectral type of the Herbig star rather than on its
mass, even though the latter is the relevant physical parameter. 
The main reasons
for this choice are the proximity of the Herbig star to the main sequence
(and hence a close relation between spectral type and mass is ``almost'' well
defined) and the much larger uncertainties in the mass estimates based
on the location of the stars in the HR diagram and the use of 
evolutionary tracks.
On the other hand, spectral types of young stars
sample are usually determined with an uncertainty of one or two
subclasses, with the exception of a few cases (see the discussion 
in Paper~II).
In particular, RNO~1B and MWC~300 do not have a reliable
classification and in the literature they are generically referred to 
as Be stars. For plotting purposes, we have arbitrarily assigned 
a B5 classification to both of them. 
These stars have both values of \Ic\ and \Nk\ typical of B stars,
and the assigned spectral type of B5 does not affect
our results.

\begin{figure}[t]
\centerline{\psfig{figure=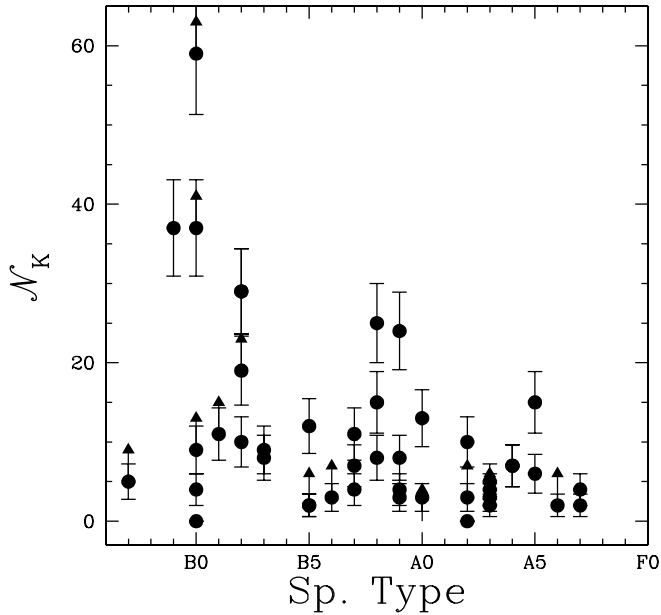,width=8.8cm}}
\caption[]{${\cal N}_{\rm K}$ as a function of the spectral type
of the Herbig Ae/Be star.
The two stars with an uncertain spectral type Be have been plotted as B5.}
\label{fncksp}
\end{figure}

\begin{figure}[t]
\centerline{\psfig{figure=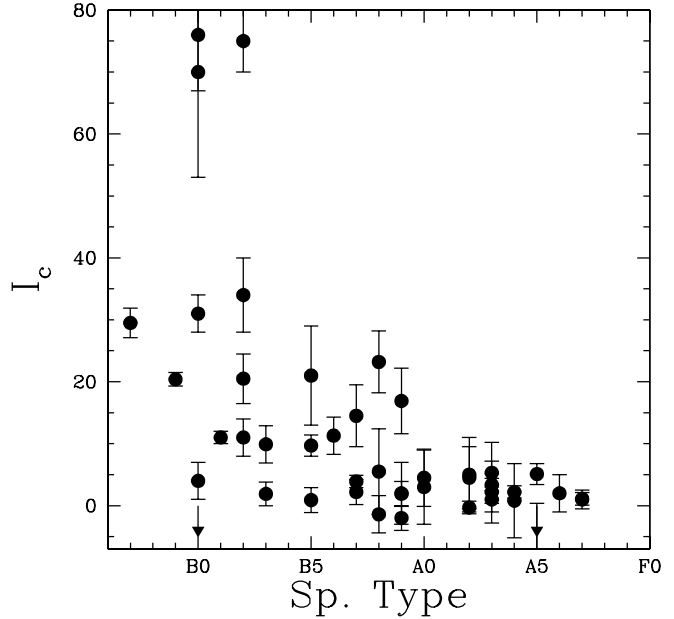,width=8.8cm}}
\caption[]{I$_{\rm C}$ versus spectral type for the whole sample.
As in Fig.~\ref{fncksp}, the two stars with an uncertain spectral type Be have been plotted as B5.}
\label{ficsp}
\end{figure}

The variation of \Nk\ and I$_{\rm C}$ as a function of the
spectral type of the Herbig star is shown in Figures~\ref{fncksp} and
\ref{ficsp}, respectively. 
Both figures confirm the initial results found in Paper~I and by 
Hillenbrand~\cite{H95} that higher mass stars tend to be surrounded by 
richer clusters.
As expected, the evidence for clusters is more pronounced using I$_{\rm C}$
instead of ${\cal N}_{\rm K}$, but the fact that the same trend is 
seen in both indicators gives strength to its real significance.

In addition to the evidence for a variation of I$_{\rm C}$ with spectral
type, the results of Fig.~\ref{ficsp} suggest the existence of three regimes 
for the distribution of stars around Herbig stars. In the first one, 
characterized by I$_{\rm C}\gsim$40, the Herbig stars are definitely
associated with rich clusters; in the intermediate regime with $10\lsim$ 
I$_{\rm C}\lsim 40$
a small cluster may be present; in the third case
where I$_{\rm C}\lsim$10 only small aggregates or background 
stars in the field are found, a situation similar to that observed 
in low-mass star forming regions.
Three stars of our sample (MWC~137, AS~310 and BD$+$65$^\circ$1637) 
belong to the first group.
The sizes of the clusters derived for these stars in Paper~II are on
the large side of the distribution ($\sim 0.4$~pc), however, at least in
two cases (AS~310 and BD$+$65$^\circ$1637) the Herbig star is not at the
center of the cluster, and this results in an overestimate of the cluster
radius, determined from the radial density profile.
Among stars of early spectral types, we find a large spread of values
of \Ic. In some cases, the low \Ic\ are dubious. For example, 
V645 Cyg and MWC~1080, have reletively low values of 
\Ic\ typical of the intermediate regime. However, both stars are embedded
in bright nebulosities which may affect the star count resulting in
a severe underestimate of the actual stellar density. In other cases, however,
the low values of \Ic\ may be real.

The presence of groups or small clusters is secure around most
of the stars belonging to the intermediate regime.
A clear central density peak is observed in the density profiles 
of almost all stars of this group, as shown in Paper~II. 
Typically, the Herbig star is located at the center of the 
stellar group, which generally has a round shape. A possible 
exception is the VY~Mon field, where the stellar aggregate has an elongated
structure with an aspect ratio of approximately 4:1 (the Herbig star is at
the center). HD~52721 is not associated with molecular gas 
(Fuente et al.~\cite{Fea98}), it is a rather old star, and the large
cluster radius is probably the result of dynamical relaxation.

Finally, the large majority ($\sim 65$\%) of the Herbig stars of the sample
have very low values of \Ic\, showing no enhancements above the background 
stellar density. All the fields around stars with spectral type later than
B9 belong to this group. However, although most of these stars have spectral
types later than B7-B8, there are extreme cases of early-type stars that 
deserve some discussion. First, the negative value of \Ic\ found for 
R~Mon (B0) is probably due to localized extinction around the star, which
is probably on the observer side of a molecular clump. Also,
the stars BHJ~71 (B0), HD~200775 (B3) and HD~259431 (B5) have anomalously 
low values of \Ic. The first one is a poorly studied star, and we have not
been able to find any information on the molecular gas in the literature
(see Paper~II). HD~200775 is in many respect very similar to
BD~65$^\circ$1637: same spectral type, same molecular gas morphology 
and similar evolutionary status (Fuente et al.~\cite{Fea98}).
The absence of a cluster of young stars around HD~200775, as opposed to the
rich cluster around BD~65$^\circ$1637, may be the result of dynamical
dissipation (see below) or reflect a difference in the environment.
HD~259431 appears to be isolated in spite of the large amount of molecular
material around it (Hillenbrand~\cite{H95}).  Among late-type stars,
LkH$\alpha$~198 stands out for its negative value of \Ic\ due to the bright
nebulosity and localized extinction. Interesting cases with an indication of
an extended population of embedded sources but not spatially concentrated
around the Herbig star are VX~Cas (A0), HD~245185 (A2) and Mac~H12 (A5).

\subsection{Mass sensitivity correction}
\label{sms}

\begin{figure}
\centerline{\psfig{figure=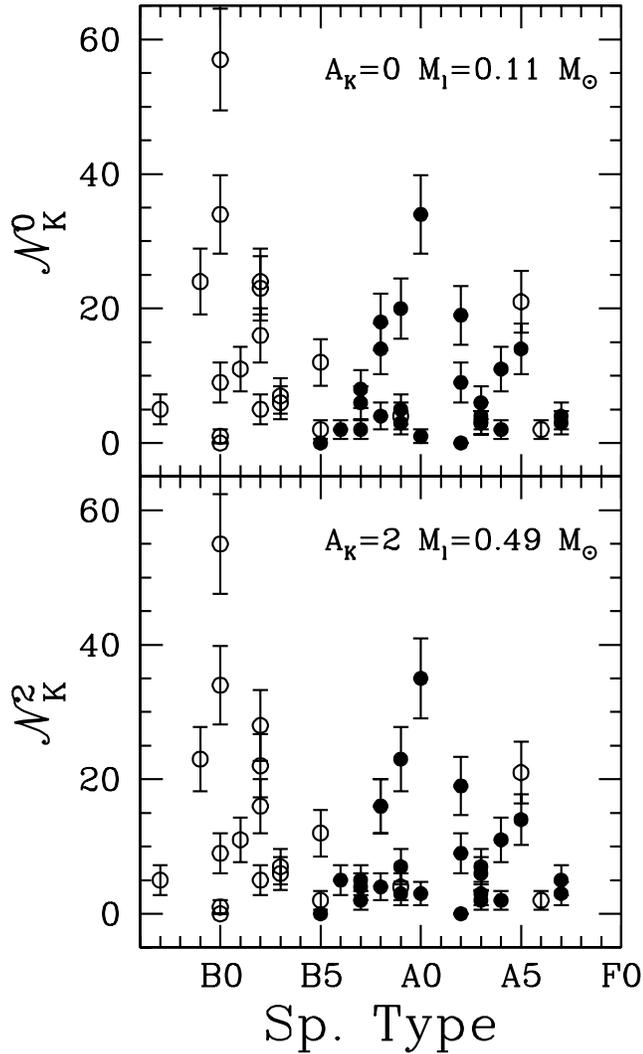,width=8.8cm}}
\caption[]{${\cal N}^0_{\rm K}$ and ${\cal N}^2_{\rm K}$
as a function of the spectral type of the Herbig Ae/Be star.
Open circles denote fields for which we could not compute the
age using the Herbig~Ae/Be star parameters (see Paper~II and sect.~\ref{sms}
of the main text).  As in Fig.~\ref{fncksp}, the
two stars with an uncertain spectral type Be have been plotted as B5.}
\label{fnckspm}
\end{figure}

\begin{figure}
\centerline{\psfig{figure=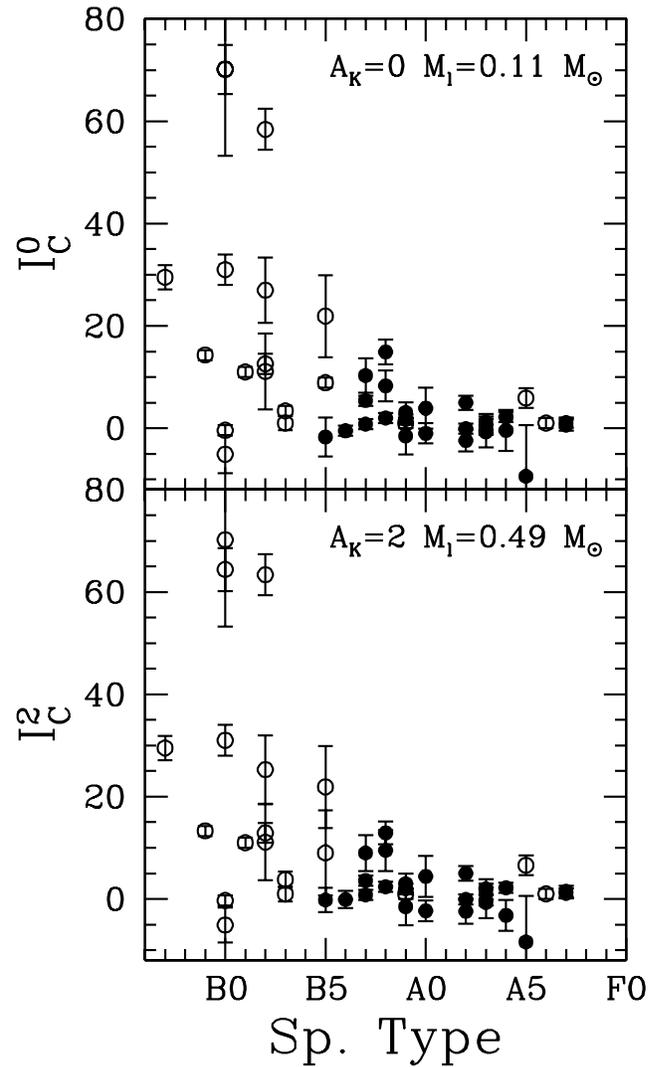,width=8.8cm}}
\caption[]{I$^0_{\rm C}$ and I$^2_{\rm C}$ versus spectral type for the
whole sample. Open circles as in Fig.~\ref{fnckspm}.
As in Fig.~\ref{fncksp}, the
two stars with an uncertain spectral type Be have been plotted as B5.}
\label{ficspm}
\end{figure}


Since young stars change their bolometric luminosity and
effective temperature during PMS evolution, the mass of
the smallest star detectable ($M_l$) in our K-band images is a function of 
age ($t_c$), of the absolute completeness K-magnitude (M$^c_K$) and of
the extinction along the line of sight (A$_K$).

In Paper~II, using the PMS evolutionary tracks for intermediate mass stars of
Palla \& Stahler~(\cite{PS93}) and the compilation of stellar parameters from
the literature, we have computed in an homogeneous way the ages of most of the
stars later than B5. Then, using the PMS evolutionary tracks
for low mass stars of D'Antona \& Mazzitelli~(\cite{DM94}), we have translated 
M$^c_K$ into $M_l$ for two values of the extinction (A$_K=0$ and 2)
assuming that all the stars in each group are 
coheval with the Herbig Ae/Be star.

There are some stars (mostly early Be types) for which it is not
possible to derive ages from PMS tracks. For these stars, we have estimated
the age in the following way. From the results of Paper~II we see that
the Herbig Ae systems tend to be 1 to 10~Myr old (the mean value is
$\sim 4.6$~Myr), while Be systems tend to be younger than 1~Myr
(with a mean age of $\sim 0.7$~Myr).
Thus we decided to adopt as age $t_c=0.5$~Myr for the 
17 Be systems and $t_c=5$~Myr for the 2 Ae systems
without an age determination in Paper~II.

If $A_K\sim$0 mag, all the fields with a good age estimate 
from Paper~II are complete to less
than 0.1 M$_\odot$, with the only exception of the two fields
around HK Ori and LkH$\alpha$198. In this case, \Nk\ and $I_c$
sample the totality of the stellar population in practically all fields.
If $A_K\sim$ 2 mag, the minimum mass is generally larger, ranging from
$<0.1$ to 0.49 M$_\odot$ in HK~Ori.
For each field we have computed the absolute magnitude corresponding 
to these masses and the corresponding extinctions at K-band, then 
we have calculated the two {\it richness indicators} discussed in 
Sect.~\ref{sres} considering, in each field, only the stars with magnitude
lower than the computed one. In this way, under the assumption that the 
mean extinction is approximately the same in all fields (and for all stars
in each field),
we have calculated mass limited {\it richness indicators}.
The values of ${\cal N}_{\rm K}$ and I$_{\rm C}$ for A$_K=0$ and 2
(${\cal N}^0_{\rm K}$,${\cal N}^2_{\rm K}$, I$^0_{\rm C}$ and I$^2_{\rm C}$)
are reported in Table~\ref{trind}, Columns~6--9.  Since some of the 
fields without an estimate of the age of the Herbig~Ae/Be star
have mass sensitivities lower than our limits 
(which are based on HK~Ori),
some of the values reported are lower limits (indicated by $>$).

In Fig.~\ref{fnckspm} and~\ref{ficspm} we show the behaviour of
the ``mass sensitivity corrected'' indicators.
Fig.~\ref{fnckspm} clearly shows that trend detected in Fig.~\ref{fncksp}
is completely cancelled by the bias introduced by the contamination from
field stars. In fact, since the late type stars are older than early type
the absolute magnitude corresponding to the same mass limit is 
higher for the Ae systems than for the Be systems, and since this discrepancy 
is not compensated by the difference in distance, we are effectively 
counting more field objects in Ae fields than in
Be fields. On the contrary, Fig.~\ref{ficspm} clearly shows the
same trend as Fig.~\ref{ficsp} supporting our conclusion of a dependence
of \Ic\ on the spectral type of the Herbig Ae/Be star. Therefore in the
following we will use \Ic\ which has a higher signal to noise ratio
than I$^0_{\rm C}$ and I$^2_{\rm C}$.

\section{Discussion}
\label{sdis}

In Paper~I we suggested the presence of a threshold 
for the appearence of clusters around Herbig stars of spectral type
earlier than $\sim$B7. 
The results discussed in the previous section
confirm the property that Be systems are
associated with more conspicuous groups than the Ae systems. Although
the identification of a critical spectral type (or mass) is no longer evident
in the complete sample, it is worth noting that no stars later 
than B9 show large groups, i.e. their I$_{\rm C}$ is $\lsim 10$.

This general result raises some important issues: (1) is this difference
between Be and Ae system related to a different location in the 
galaxy? (2) since Ae systems are generally older than Be systems,
could the disappearence of the cluster be an evolutionary effect?
(3) what is the typical environment in which intermediate mass
stars form? (4) does the standard accretion scenario for the formation
of Herbig stars need to be replaced by formation in clusters?

The first question is easy to answer.
We do not find evidence for a dependence of
the clustering properties with the galactic position of the target star,
shown in Paper~II.
We must caution, however, that most of the observed stars 
lie in the outer regions of the Galaxy, due to the selection effect
introduced by our observing sites in the northern hemisphere. 
Although we do not expect an opposite result toward the inner Galaxy,
we must await the results of a similar study on a southern sample
before generalizing our conclusion.

\subsection{Age effect}
\label{sage}

Following the age estimates given in Paper~II, our sample of Herbig~Ae/Be
stars covers the range of
ages from $\sim 0.05$ to 10~Myr. As noted above, Ae stars tend to be 
older than Be stars. Thus, it is possible that the variation of the
{\it richness indicators} from Be to Ae systems is caused by 
some evolutionary effect. It is possible that a stellar 
group composed by few tens or hundreds of objects could be dispersed on 
a timescale of a few million years and become
undetectable (i.e. confused with the field stars) in older systems.

\begin{figure}
\centerline{\psfig{figure=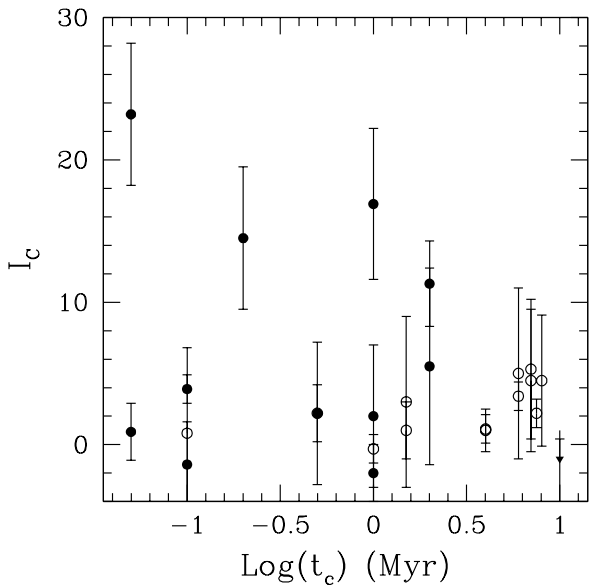,width=8.8cm}}
\caption[]{I$_{\rm C}$ versus age.
Only the fields with age determination from Paper~II are presented.
The arrow represents the field around
LkH$\alpha$~198 (I$_{\rm C}=-10.6$).
filled circles: Be systems; open circles: Ae systems.}
\label{fage}
\end{figure}

In order to explore this possibility more quantitatively, 
we show in Fig.~\ref{fage} the run of I$_{\rm C}$ as a function of the age
for the 25 fields with an age determination of the
Herbig star (see Table~2 of paper~II).
The lack
of stars with high values of I$_{\rm C}$ for ages greater than 2~Myr
is related to the fact that our sample does not contain Be stars
of that age. However, it is worth noting that for Be systems
(filled circles in Fig.~\ref{fage}) younger
than 2~Myr we find high and low values of I$_{\rm C}$ at any age.

From the figure we see that it does not seem to exist a correlation between
the age of the Herbig star and the presence of a cluster. The correlation
is only with the spectral type (or mass). Stars earlier than B7 have 
clusters or not independently of their age; stars later than B7 do not have 
clusters and are on average older than the rest. If we say for convenience 
that a cluster is detected for $I_{\rm C}\gsim 10$, the oldest star
surrounded by a cluster is
XY~Per with an age of 2 Myr. Since stars of age greater than $\sim$4 Myr 
do not have clusters, there are two possibilities to account for such 
property: these stars did form in relative isolation or the attending 
cluster has by now disappeared. The former explanation is obvious and 
does not require any further comment.

The second alternative is theoretically more interesting and
we have performed several N-body simulations to verify under which 
conditions the dynamical evolution of a small cluster can evolve so rapidly
to dissipate the majority of its members in a 
time scale less than $\sim$4~Myr.
We have considered models
containing $N=$100, 150 and 200 stars starting
from different initial conditions (virialized and non virialized clusters) 
and with varying frequency distributions of stellar masses (Salpeter, Scalo, 
Kroupa and uniform mass). The half-mass radius was varied between 0.4 and
0.7 pc and the typical velocity dispersion was 2 km s$^{-1}$. 
Thus, the clusters are initially richer and larger than those actually
observed around Herbig stars characterized by I$_{\rm C}\sim$30-40 and
a radius of $\sim$0.2 pc.
We have also
varied the upper limit of the most massive object from 20 M$_\odot$,
corresponding to a O9 star, to 2 M$_\odot$, corresponding to an A5 star.
It is well known that the dynamical evolution of cluster models is very
dependent on the choice of the IMF, especially in what concerns the
effects of mass segregation at early times (e.g. de la Fuente Marcos 1995, 
1997). We have not considered the presence of primordial binaries in
the initial population. 

For most runs, we have found that it is very difficult
to loose a significant fraction of stars in the initial few crossing times,
as required by the age constraints of the Herbig stars. The only cases where
the stellar population decreases to less than half the initial value
in about ten crossing times are those characterized
by the lowest number of stars ($N=$100 in our models) distributed in
mass according to a Salpeter IMF. 
The first point is a critical one, since
the evaporation time increases very rapidly with the number of stars.
Even models with $N=$150 do not evolve sufficiently rapidly. As for
the IMF, the result is acceptable even though we know that a single power-law
approximation must break down at subsolar values. More important, however,
is the sensitivity of the evolution to the upper limit of the mass
distribution. Only if the most massive star exceeds 5~M$_\odot$, the evolution
is fast enough that the system loses many members in several crossing times. 
Otherwise, it is impossible to modify the cluster composition in a few 
million years.
Now, since a 5~M$_\odot$ corresponds to a B6 ZAMS star, it is clear that
Herbig stars of later types which were formed in relatively small clusters
would have retained the original population of companions at the observed
age. {\it Therefore, we conclude that the fact that we do not see evidence
for the presence of clusters around stars of type later than about B7-B8
is an imprint of the stellar formation mode rather than the consequence
of the dynamical evolution of a presently dispersed cluster. On the other
hand, the absence of clusters around some of the early type B stars could
result from
the dynamical evolution, especially in case of a relatively small initial
population}.

\subsection{The transition from loose to dense groups}
\label{srho}

\begin{figure*}
\centerline{\psfig{figure=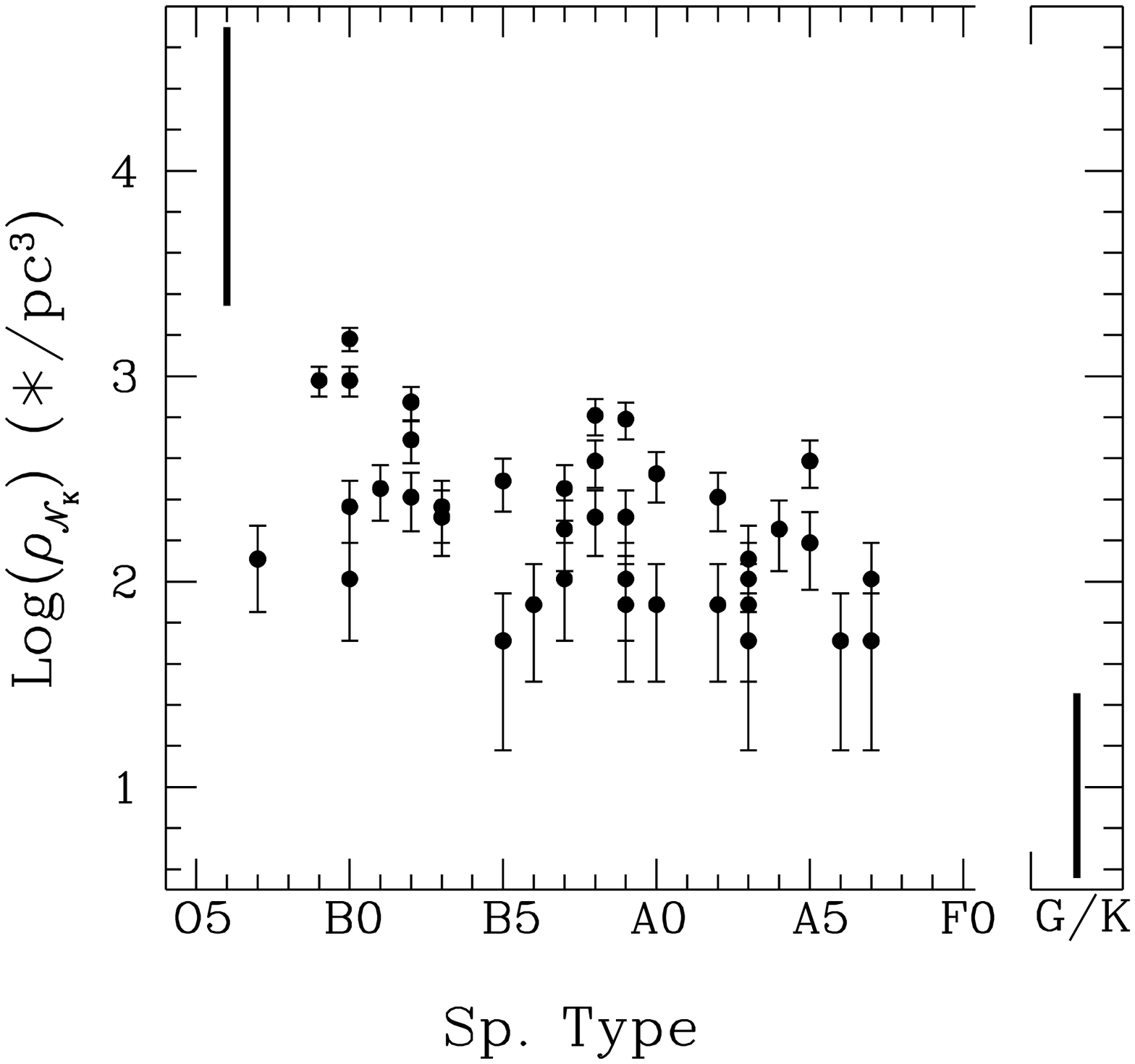,height=8cm}
	    \hskip 0.3cm
	    \psfig{figure=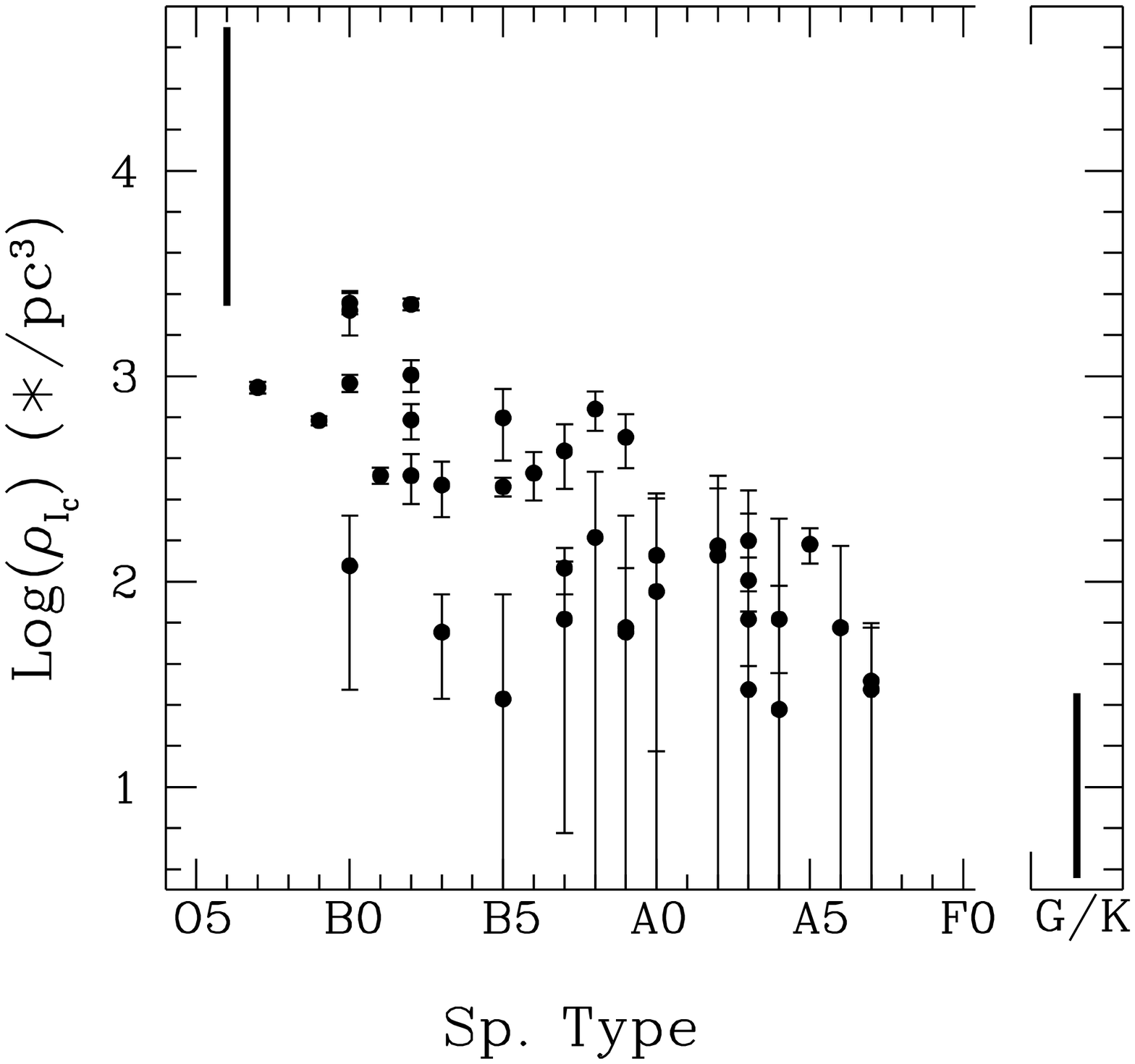,height=8cm}}
\caption[]{\label{frho} Stellar volume densities derived
from ${\cal N}_{\rm K}$ ({\it left}) and from I$_{\rm C}$ ({\it right})
versus spectral type of the central star. 
Stars with I$_{\rm C}<0$ have been excluded. 
The heavy vertical line at O6 represents the range of
stellar densities found in the Trapezium cluster,
whereas that at G/K (not to scale) represents
the densities of stellar groups in Taurus-Auriga.}
\end{figure*}

The values of ${\cal N}_{\rm K}$ and I$_{\rm C}$ can be transformed
into star (or effective star) volume densities using the average 
cluster radius of 0.2 pc and assuming spherical clusters. The logarithm
of the stellar densities in sources per cubic parsec are reported in the last
two columns of Table~\ref{trind}. 
This estimates of the stellar densities are affected by large uncertainties
(as high as 30\%, Hillenbrand~\cite{H95}),
due to the unknown depth of the clusters along the line of sight and the 
neglection of concentration effects, but it is a useful approximation 
to be used for comparing the various regions.
The resulting distributions shown in Fig.~\ref{frho} confirm the apparent
trend discussed in Sect.~2 and put on a firmer ground the suggestion
of Hillenbrand (1995) of a physical relationship between cluster
density and the maximum mass of the Herbig star.
For comparison, we also show in
Fig.~\ref{frho} the stellar density range of the T Tauri aggregates found in
Taurus-Auriga by Gomez et al.~(\cite{Gea93}) (shown
as the solid line at G/K, not to scale),
and the same quantity for the 
Trapezium cluster (McCaughrean \& Stauffer~\cite{McCS94}; 
Hillenbrand~\cite{H97}). The conclusion is evident:
{\it intermediate mass stars mark the 
transition from low density aggregates of $\lsim 10$ stars per cubic parsec
of T Tauri stars to dense clusters of $\gsim 10^3$ stars per cubic 
parsec associated with early-type stars}.

\subsection{Implications for star formation}
\label{sisf}

The conventional theory of protostellar infall successfully accounts
for the formation and early stellar evolution of low- and intermediate-mass 
stars up to about 10 M$_\odot$. The location of the stellar birthline 
and the distribution of the observed T Tauri and Herbig Ae/Be stars in
the HR diagram are in excellent agreement with the theoretical 
predictions (Palla \& Stahler 1990, 1993). Protostars more massive than 
about 10 M$_\odot$ burn hydrogen while still in the accretion phase and 
therefore join the main sequence, implying that stars of even higher 
mass have no 
contracting PMS phase. Observations of clusters such as NGC~6611 and the
dense regions of the Trapezium cluster support this finding (e.g. Hillenbrand
1993, 1998). However, the accretion scenario of isolated protostars
fails to explain the existence of massive stars. Radiation pressure by
photons produced at the stellar and disk surfaces on the infalling gas begins
to become significant at about the critical mass of 10-15 M$_\odot$ 
(e.g. Larson \& Starrfield 1974; Yorke \& Krugel 1977). This limit can be 
increased by considering variations in the dust properties (abundances and 
sizes) or in the mass accretion rate (Wolfire \& Cassinelli 1987). But in 
either case the required
conditions are so extreme (dust depletion by at least one order of 
magnitude and accretion rates of at least 10$^{-3}$~M$_\odot$~yr$^{-1}$)
that cannot reasonably apply in all circumstances. Any departure of the
infall from spherical symmetry can also help to shift the critical mass
to very massive objects (e.g. Nakano 1989; Jijina \& Adams 1996).

The fact that, as we have seen, stars form in groups and clusters rather
than in isolation may offer an alternate mechanism to circumvent the 
problem of the high luminosity of massive stars and the negative feedback on
the accreting gas. Accretion processes in clusters has long been advocated to
explain the shape of the initial mass function (Zinnecker 1982; Larson 1992;
Price and Podsiadlowski 1995), and to account for the location of massive
stars in the cluster centers (Bonnell et al. 1997). Very recently, Bonnell et
al. (1998) have presented model calculations of the dynamical evolution of
rich, young clusters that show that accretion-induced collisions of low- and
intermediate-mass stars (formed in the standard accretion mode) can result in
the formation of more massive objects in time scales less than 10$^6$ years.
However, the process requires a critical stellar/protostellar density so that
collisions and merging are frequent. For the formation of a star of mass of
$\sim$50 M$_\odot$, an initial density of $\rho_\ast>$10$^4$ stars pc$^{-3}$
is required. This is the typical density observed in the central regions of
large clusters such as the Orion Nebula cluster where stars so massive are
indeed present, but it is hardly the case for the stars of our sample. As
shown in Fig.~\ref{frho}, the highest stellar volume densities are limited to
about 2$\times$10$^3$ stars pc$^{-3}$, almost an order of magnitude lower
than the minimum value for the efficient collisional build-up of massive
stars. These densities are found for stars of spectral type B0-B2, i.e. with
ZAMS masses of about 20 to 10 M$_\odot$. Thus, these stars are at the
borderline between the conditions of isolated and collisional accretion
mechanisms.  It would be extremely interesting to extend the Bonnell et al.
(1998) models to less extreme conditions on the stellar density in order to
explore the dynamical evolution of clusters resembling those observed around
Herbig Be stars.  Based on the observational evidence, we may conclude that
{\it in most cases, the formation of Herbig stars can be understood in terms
of the conventional accretion scenario, whereas for the most massive members
of this group dynamical interactions in the cluster core and residual gas
accretion can be of importance in determining the final mass, even though the
observed stellar density is never extremely high}.

\section{Conclusions}
\label{scon}

Using the techniques described in Paper~I,
we have analyzed the near-infrared observations 
of the fields surrounding Herbig Ae/Be stars presented in Paper~II
with the goal of identifying and characterizing the presence of (partially)
embedded clusters formed around the intermediate-mass stars.
We have examined 44 fields around stars ranging in spectral types
from A7 to O9. The main results of these studies
can be summarized as follows.

We have confirmed and extended the correlation between the spectral
type (or mass) of the Herbig stars and the number of nearby, lower mass
objects, first noted by Hillenbrand (1995). Rich clusters with 
densities up to 10$^3$ pc$^{-3}$ only appear around stars earlier
than B5, corresponding to a mass of about 6 M$_\odot$. Conversely,
A-type stars are never accompanied by conspicuous groups and the
typical density is less than 50 pc$^{-3}$. However,
the transition between formation in loose groups 
and in clusters does not occur sharply around spectral type B7 as
we suggested in Paper~I. The appearence of denser stellar groups 
is quite smooth moving from Ae to Be systems, thus suggesting
that intermediate-mass stars fill naturally the gap between 
the low-density, low-mass aggregates and the high-density, high-mass clusters.

Using a richness indicator, I$_{\rm C}$, based on stellar 
surface density profiles,
we have identified three regimes for the distribution of stars
around Herbig Ae/Be stars: rich clusters characterized by values
of I$_{\rm C}\gsim$40; small clusters or aggregates with
$10\lsim$I$_{\rm C}\lsim 40$,
and small aggregates or background stars for lower values of I$_{\rm C}$.
Herbig stars with rich clusters are rare:
only three stars of the sample definitely belong to the first group,
whereas in other cases we have indication that the star counts
may severely underestimate the actual stellar density.
The typical cluster size is $\sim$0.2 pc irrespective of the richness
of the clusters. This value is remarkably close to the dimension
of dense cores in molecular clouds.

The majority of Herbig stars (65\%) are found in the last regime,
showing no enhancement above the background stellar density.
All the stars with spectral types later than B9 belong to this regime
and are generally
older than Be stars. This result is not affected
by different sensitivities to the lowest masses in different
fields. From dynamical considerations,
we conclude that the absence of clusters around late-Be and Ae stars is an
imprint of stellar formation in relatively isolated cores (as in the case
of T Tauri stars), and not the result of rapid cluster evolution and 
dispersion.

The fact that the most massive Herbig stars have clusters strongly supports
the notion that their formation has been influenced by dynamical
interactions with lower mass stars and/or protostellar cores.
It is no coincidence that the onset of clusters manifests at a
mass of about 8-10 M$_\odot$ where the conventional accretion scenario
for isolated protostars faces severe theoretical problems.
Future studies of the formation and evolution of these stars should
take into account the observational evidence for high stellar density
environments.

\begin{acknowledgements}
It is a pleasure to thank Alessandro Navarrini for his help with the
N-body simulations described in this paper.
This project was partly supported by ASI grant ARS 96-66 and by CNR grant
97.00018.CT02 to the Osservatorio di Arcetri.
Support from CNR-NATO Advanced Fellowship program and 
from NASA's {\it Origins of Solar Systems} program (through grant NAGW--4030)
is gratefully acknowledged.
\end{acknowledgements}

\end{document}